\documentclass[aps,prb,twocolumn,superscriptaddress,floatfix]{revtex4}

\usepackage[dvipdfmx]{graphicx}
\usepackage[dvipdfmx]{color}
\usepackage{amsmath}
\usepackage{amssymb}
\usepackage{dcolumn}
\usepackage{bm}
\usepackage{latexsym} 
\usepackage{setspace}
\usepackage{graphicx}
\usepackage{color}
\usepackage[dvipdfmx,colorlinks=true,linkcolor=blue,citecolor=blue]{hyperref}
\begin{document}
\title{Composition-tunable magnon-polaron anomalies in spin Seebeck effects \\ in epitaxial Bi$_x$Y$_{3-x}$Fe$_{5}$O$_{12}$ films} 
\author{Takashi Kikkawa}
\email{t.kikkawa@ap.t.u-tokyo.ac.jp}
\affiliation{Department of Applied Physics, The University of Tokyo, Tokyo 113-8656, Japan}
\affiliation{WPI Advanced Institute for Materials Research, Tohoku University, Sendai 980-8577, Japan}
\affiliation{Institute for Materials Research, Tohoku University, Sendai 980-8577, Japan}
\author{Koichi Oyanagi}
\affiliation{Institute for Materials Research, Tohoku University, Sendai 980-8577, Japan} 
\affiliation{Faculty of Science and Engineering, Iwate University, Morioka 020-8551, Japan}
\author{Tomosato Hioki}
\affiliation{WPI Advanced Institute for Materials Research, Tohoku University, Sendai 980-8577, Japan}
\affiliation{Department of Applied Physics, The University of Tokyo, Tokyo 113-8656, Japan} 
\author{Masahiko Ishida}
\affiliation{Secure System Platform Research Laboratories, NEC Corporation, Kawasaki 211-8666, Japan}
\author{Zhiyong Qiu}
\affiliation{Institute for Materials Research, Tohoku University, Sendai 980-8577, Japan}
\affiliation{School of Materials Science and Engineering, Dalian University of Technology, Dalian 116024, China}
\author{Rafael Ramos} 
\affiliation{WPI Advanced Institute for Materials Research, Tohoku University, Sendai 980-8577, Japan} 
\affiliation{Centro de Investigaci\'{o}n en Qu\'{i}mica Biol\'{o}xica e Materiais Moleculares (CIQUS), Departamento de Qu\'{i}mica-F\'{i}sica, Universidade de Santiago de Compostela, Santiago de Compostela 15782, Spain}
\author{Yusuke Hashimoto} 
\affiliation{WPI Advanced Institute for Materials Research, Tohoku University, Sendai 980-8577, Japan} 
\author{Eiji Saitoh}
\affiliation{Department of Applied Physics, The University of Tokyo, Tokyo 113-8656, Japan}
\affiliation{WPI Advanced Institute for Materials Research, Tohoku University, Sendai 980-8577, Japan}
\affiliation{Institute for AI and Beyond, The University of Tokyo, Tokyo 113-8656, Japan}
\affiliation{Advanced Science Research Center, Japan Atomic Energy Agency, Tokai 319-1195, Japan}
\date{\today}
\begin{abstract} 
We have investigated hybridized magnon-phonon excitation (magnon polarons) in spin Seebeck effects (SSEs) in Bi$_x$Y$_{3-x}$Fe$_{5}$O$_{12}$ (Bi$_x$Y$_{3-x}$IG; $x=0$, $0.5$, and $0.9$) films with Pt contact. 
We observed sharp peak structures in the magnetic field $H$ dependence of the longitudinal SSE (LSSE) voltages, which appear when the phonon
dispersions are tangential to the magnon dispersion curve in Bi$_x$Y$_{3-x}$IG.  
By increasing the Bi amount $x$, the peak fields in the LSSE shift toward lower $H$ values due to the reduction of the sound velocities in Bi$_x$Y$_{3-x}$IG.  
We also measured the SSE in a nonlocal configuration and found that magnon-polaron anomalies appear with different signs and intensities.  
Our result shows composition-tunability of magnon-polaron anomalies and provides a clue to further unravel the physics of magnon-polaron SSEs.  
\end{abstract} 
\maketitle
%
%
%
\section{INTRODUCTION} \label{sec:introduction}
\begin{figure}[tbh]
\begin{center}
\includegraphics{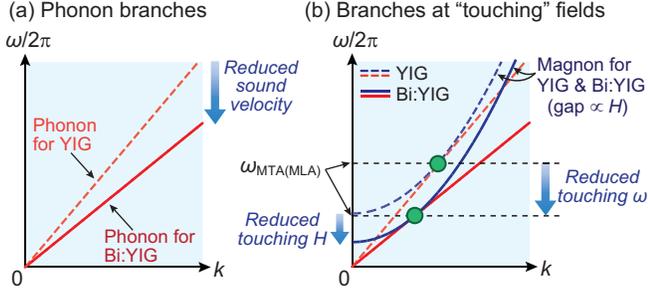}
\caption{ 
(a) A schematic illustration of the phonon dispersion relations for pure YIG and Bi-substituted YIG (Bi:YIG).
(b) A schematic illustration of the magnon and phonon dispersion relations for pure YIG and Bi:YIG at the touching field. 
Owing to the reduction of the sound velocity by the Bi substitution, the touching field $H_{\rm TA(LA)}$ and the touching angular frequency $\omega_{\rm MTA(MLA)}$ between the magnon and TA(LA)-phonon branches for Bi:YIG shift toward lower values compared to those for YIG.  
}
\label{fig:BIYIG-dispersions-schematics}
\end{center}
\end{figure}

%
Bismuth (Bi) substituted yttrium iron garnet (YIG) has been an important material in magnetics and magneto-optics \cite{Hansen1983PRB,Hansen1984ThinSolidFilms}. 
In spite of Bi$^{3+}$ being a diamagnetic ion, via the spin-orbit interaction, the substitution dramatically increases the Faraday rotation angle of garnets (by a factor of $\sim 10^2$ compared to primitive YIG \cite{Hansen1983PRB,Hansen1984ThinSolidFilms}), which led to its versatile magneto-optical application. 
Another interesting feature of Bi-substituted YIG (Bi:YIG) includes the increased Curie temperature \cite{Hansen1983PRB,Hansen1984ThinSolidFilms} and magnetoelastic interaction \cite{Kumar2019JPCM}. Fabrication of strained Bi:YIG ultrathin films with perpendicular magnetic anisotropy and low Gilbert damping has also been reported, offering a new opportunity for their spintronic applications \cite{Soumah2018NatCommun,Evelt2018PRAppl,Chumak2022IEEE}. 
Furthermore, because of the atomic mass of Bi ($= 209$) much larger than that of Y ($= 89$), a decrease of the sound velocities of Bi:YIG ($c_{\rm TA, LA}^{\rm BiYIG}$) has been reported for both the transverse acoustic (TA) and longitudinal acoustic (LA) modes compared to those of YIG ($c_{\rm TA, LA}^{\rm YIG}$) \cite{Zhang1993PRB,Siu2001PRB}, as schematically shown in Fig. \ref{fig:BIYIG-dispersions-schematics}(a) (see also Table \ref{tab:comparison}) . \par 
Lately, in spintronics, magnon-phonon coupled phenomena have renewed attention, where not only magnon but also phonon dispersion relations play an important role \cite{Dreher2012PRB,Ogawa2015PNAS,Shen2015PRL,Guerreiro2015PRB,Kikkawa2016PRL,Bozhko2017PRL,Hashimoto2017NatCommun,Holanda2018NatPhys,Hashimoto2018PRB,Hayashi2018PRL,Yahiro2020PRB,An2020PRB,Godejohann2020PRB,Hioki2020ComPhys,Frey2021PRB,Zhang2021NatCommun,Schlitz2022PRB}.  
Of particular interest is the region of the crossings of their branches, at which magnons and phonons are allowed to hybridize by the magnetoelastic interaction, giving rise to the magnon-polaron modes having both magnonic and phononic characters \cite{Shen2015PRL,Kikkawa2016PRL,Flebus2017PRB}. 
In thermal magnon-spin transport, or the spin Seebeck effect (SSE) \cite{Uchida2010APL,Uchida2014JPCM,Uchida2010ProcIEEE,Kikkawa2023ARCMP}, the magnon-polaron formation manifests as anomalous peak or dip structures at the onset fields $H_{\rm TA,LA}$, where the phonon dispersions in a magnet tangentially touch the magnon dispersion gapped by the Zeeman interaction ($\propto$ external magnetic field $H$) [see Fig. \ref{fig:BIYIG-dispersions-schematics}(b)] \cite{Kikkawa2016PRL,Flebus2017PRB}.   
A calculation based on Boltzmann transport theory reveals that, when the scattering rate of magnons $\tau^{-1}_{\rm mag}$ is larger (smaller) than that of phonons $\tau^{-1}_{\rm ph}$, magnon polarons may have a longer (shorter) relaxation time than pure magnons. The SSE intensity is therefore enhanced (suppressed) at the touching fields, where the effect of magnon-polaron formation is maximal \cite{Flebus2017PRB}.   
So far, through longitudinal SSE (LSSE) experiments, magnon-polaron peaks have been detected for several magnetic films, indicating the situation of $\tau^{-1}_{\rm mag} > \tau^{-1}_{\rm ph}$ \cite{Kikkawa2016PRL,Wang2018APL,Ramos2019NatCommun,Xing2020PRB,Li2020PRL_Cr2O3,Yang2021PRB}, while dips were observed for some specific YIG bulk samples (i.e., $\tau^{-1}_{\rm mag} < \tau^{-1}_{\rm ph}$) \cite{Kikkawa2023ARCMP,Shi2021PRL_YIG-bulk}. \par 
Here, we report SSEs in epitaxially grown Bi$_x$Y$_{3-x}$Fe$_{5}$O$_{12}$ (Bi$_x$Y$_{3-x}$IG with $x= 0$, $0.5$, and $0.9$) films. The Bi$_x$Y$_{3-x}$IG films may provide an interesting platform to systematically study magnon-polaron anomalies in SSEs, as the films exhibit the enhanced magnetoelastic coupling constant $B_{\perp}$ \cite{Kumar2019JPCM} and the reduced sound velocities $c_{\rm TA,LA}$ by increasing the Bi amount. The former feature leads to an increase in magnon-phonon hybridized region in momentum space (i.e., increase in the anticrossing gap $\propto B_{\perp}$ \cite{Flebus2017PRB}), and is thus advantageous for magnon-polaron SSEs. The latter feature decreases the magnon-phonon touching fields, since they scale with the sound-velocity squared, as discussed later [$H_{\rm TA(LA)} \propto c_{\rm TA (LA)}^2$, see Eq. (\ref{equ:touching-field})]. This situation can also be understood intuitively through a sketch of the magnon-phonon branches at the touching field; as shown in Fig. \ref{fig:BIYIG-dispersions-schematics}(b), the external $H$ required for making the touching condition decreases for Bi:YIG having small $c_{\rm TA(LA)}$ compared to YIG. 
In this paper, we first show structural and magnetic characterization of the epitaxial Bi$_x$Y$_{3-x}$IG films used in this study and evaluate their sound velocities and magnetoelastic coupling constant (from Sec. \ref{sec:XRD-TEM} to Sec. \ref{sec:MEC}). 
We then show experimental results on the LSSE in the Pt/Bi$_x$Y$_{3-x}$IG films, where decreased anomaly fields are indeed observed for Bi:YIG (Sec. \ref{sec:LSSE}). 
We also performed experiments on the SSE in a nonlocal configuration, in which magnon-polaron anomalies
appear differently from those in the longitudinal configuration in terms of their sign and intensity (Sec. \ref{sec:nlSSE}).  
Our results show composition-tunability of magnon-polaron anomalies and provide a clue to further understand magnon-polaron SSEs. \par 
\begin{figure}[tbh]
\begin{center}
\includegraphics{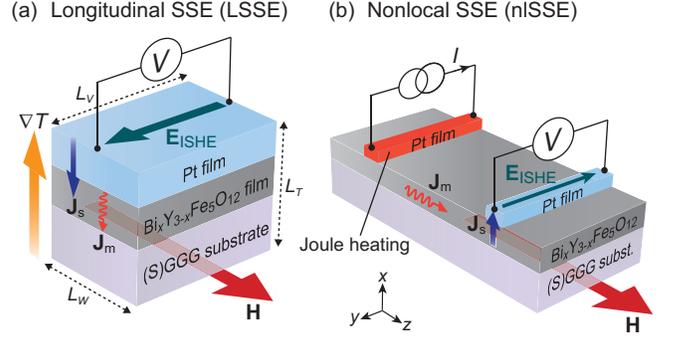}
\caption{ 
Schematic illustrations of the (a) LSSE and (b) nlSSE in Pt/Bi$_x$Y$_{3-x}$IG/(S)GGG, where $\mathbf{E}_{\mathrm{ISHE}}$, $\mathbf{J}_{\mathrm{s}}$, $\mathbf{J}_{\mathrm{m}}$, $\nabla T$, and ${\bf H}$ represent the electric
field induced by the ISHE, spin-current injection, magnon-flow direction in Bi$_x$Y$_{3-x}$IG, temperature gradient, and external magnetic field, respectively. 
$\mathbf{J}_{\mathrm{m}} \perp {\bf H}$ for the LSSE, while $\mathbf{J}_{\mathrm{m}} ~||~ {\bf H}$ for the nlSSE.
}
\label{fig:LSSE-nlSSE}
\end{center}
\end{figure}
%
\begin{figure*}[htb]
\begin{center}
\includegraphics[width=17cm]{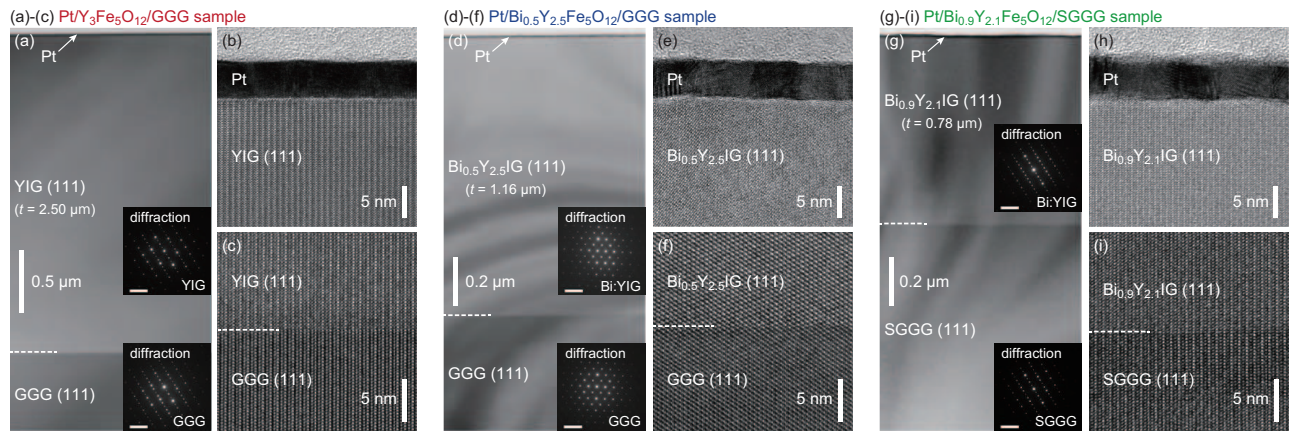}
\end{center}
\caption{Cross-sectional TEM images of the (a)-(c) Pt($5~\textrm{nm}$)/YIG/GGG, (d)-(f) Pt($5~\textrm{nm}$)/Bi$_{0.5}$Y$_{2.5}$IG/GGG, and (g)-(i) Pt($5~\textrm{nm}$)/Bi$_{0.9}$Y$_{2.1}$IG/SGGG samples.  
The TEM images shown in (a), (d), and (g) provide the overall sample cross-sections.  
The insets in (a), (d), and (g) show the selected area diffraction patterns, showing good agreement of the patterns between the (a) YIG and GGG, (d) Bi$_{0.5}$Y$_{2.5}$IG and GGG, and (g) Bi$_{0.9}$Y$_{2.1}$IG and SGGG layers. The white scale bars in the diffraction patterns represent $5~\textrm{nm}^{-1}$. 
Note that the cross sections of the TEM lamellae specimens were arbitrary crystalline planes  (i.e., the incident electron beam was not directed along specific crystalline axes), so that the different diffraction patterns appear for each sample. 
(b),(c),(e),(f),(h),(i) High resolution TEM images of the (b) Pt/YIG, (c) YIG/GGG, (e) Pt/Bi$_{0.5}$Y$_{2.5}$IG, (f) Bi$_{0.5}$Y$_{2.5}$IG/GGG, (h) Pt/Bi$_{0.9}$Y$_{2.1}$IG, and (i) Bi$_{0.9}$Y$_{2.1}$IG/SGGG interfaces.  
}
\label{fig:TEM}
\end{figure*}
%
%
%
\section{SAMPLE PREPARATION AND EXPERIMENTAL SETUP} \label{sec:procedure}
We have grown three types of Bi$_x$Y$_{3-x}$IG films by a liquid phase epitaxy (LPE) method \cite{Blank1972LPE,Simsa1984LPE,Keszei2001LPE,Kono2006LPE,Qiu2013APL,Qiu2015APEX}: YIG, Bi$_{0.5}$Y$_{2.5}$IG, and Bi$_{0.9}$Y$_{2.1}$IG with the thickness of $2.50$, $1.16$, and $0.78~\mu \textrm{m}$, respectively.
The YIG and Bi$_{0.5}$Y$_{2.5}$IG films were grown on (111) planes of 0.5-mm-thick GGG substrates, while the Bi$_{0.9}$Y$_{2.1}$IG film was grown on a (111) plane of a 0.7-mm-thick  (GdCa)$_3$(GaMgZr)$_{5}$O$_{12}$ (substituted-GGG; SGGG) substrate to reduce the lattice mismatch between the film and substrate layers. Here, the lattice constant for cubic YIG ($a_{{\rm YIG}}$), Bi$_{0.5}$Y$_{2.5}$IG ($a_{{\rm Bi}_{0.5}{\rm Y}_{2.5}{\rm IG}}$), Bi$_{0.9}$Y$_{2.1}$IG ($a_{{\rm Bi}_{0.9}{\rm Y}_{2.1}{\rm IG}}$), GGG ($a_{{\rm GGG}}$), and SGGG ($a_{{\rm SGGG}}$) are $12.376$, $12.417$, $12.450$, $12.383$, and $12.508$ ${\rm \AA}$, respectively \cite{Chern1997JJAP,Siu2001PRB}.  
The LPE fabrication for the pure YIG film ($x = 0$) was done in PbO-B$_2$O$_3$ flux at the temperature of $T = 1210~\textrm{K}$ \cite{Qiu2013APL}, while that for the Bi$_x$Y$_{3-x}$IG film with $x=0.5$ ($0.9$) was done in PbO-Bi$_2$O$_3$-B$_2$O$_3$ flux at $1053~\text{K}$ ($1023~\text{K}$).  
After the growth, the Bi$_x$Y$_{3-x}$IG films were cleaned with so-called Piranha etch solution (a mixture of H$_2$SO$_4$ and H$_2$O$_2$ at a ratio of 1:1) and also with acetone in an ultrasonic bath before Pt deposition.
The film composition was characterized by an electron probe micro analyzer in the wavelength dispersive mode. 
The crystallinity and lattice parameters of the Bi$_x$Y$_{3-x}$IG films were analyzed by means of high-resolution X-ray diffraction (XRD)  
and transmission electron microcopy (TEM), from which the film thickness was evaluated.  
The magnetic properties were measured using VSM (vibrating sample magnetometer) option of  Physical Property Measurement System (PPMS), Quantum Design Inc. \par
For LSSE measurements, we fabricated 5-nm-thick Pt films on the Bi$_x$Y$_{3-x}$IG films to electrically detect the SSE voltage based on the inverse spin-Hall effect (ISHE) \cite{ISHE_Azevedo,ISHE_Saitoh,ISHE_Valenzuela,ISHE_Costache,ISHE_Kimura}  [see Fig. \ref{fig:LSSE-nlSSE}(a)]. 
Here, the Pt films were prepared by ex-situ d.c. magnetron sputtering in a 10$^{-1}$ Pa Ar atmosphere \cite{Nozue2018APL}.
The samples were cut into a rectangular shape whose length $L_{V}$, width $ L_{W}$, and thickness $ L_{T}$ are, respectively, $\sim$ 4 mm, 2 mm, and 0.5 (0.7) mm for the Pt/YIG/GGG and Pt/Bi$_{0.5}$Y$_{2.5}$IG/GGG (Pt/Bi$_{0.9}$Y$_{2.1}$IG/SGGG) structures. 
To apply a temperature difference, $\Delta T$, the sample was sandwiched between two sapphire plates; the temperature $T_{\rm L}$ of the bottom sapphire plate in contact with the (S)GGG substrate is varied in the range from 300 to 3 K, while the temperature $T_{\rm H}$ of the top sapphire plate placed on the Pt layer is increased by an attached heater \cite{Kikkawa2015PRB,Ito2019PRB}.   
A uniform magnetic field ${\bf H}$ (with the magnitude $H$) was applied parallel to the film interface [along the $z$ direction in Fig. \ref{fig:LSSE-nlSSE}(a)] by a superconducting solenoid magnet.  
We measured the d.c. electric voltage difference $V$ between the ends of the Pt films. Hereafter, we plot the LSSE coefficient defined as $S \equiv (V/L_V)/(\Delta T/L_T)$. \par 
%
%
%
For nonlocal SSE (nlSSE) measurements, we prepared nanofabricated Pt/Bi$_x$Y$_{3-x}$IG/Pt devices, where two electrically-separated Pt strips with the distance of $d = 8~\mu\textrm{m}$ are formed on the Bi$_x$Y$_{3-x}$IG films, as schematically shown in Fig. \ref{fig:LSSE-nlSSE}(b). 
Here, Pt strips were prepared by means of electron-beam lithography, followed by Pt sputtering and a lift-off process \cite{Cornelissen2017PRB,Oyanagi2020AIPAdv}, whose dimensions are $200~\mu\textrm{m}$ length ($l_{V}$), $100~\textrm{nm}$ width, and $10~\textrm{nm}$ thickness. 
In our nlSSE setup, the Joule heating of an applied charge current ($I$) to the one Pt strip generates a magnon spin current in Bi$_x$Y$_{3-x}$IG. When some of the magnons reach the other Pt strip, they are converted into a conduction-electron spin current and subsequently detected as an ISHE voltage [see Fig. \ref{fig:LSSE-nlSSE}(b)]. 
We measured the nlSSE using a lock-in detection technique \cite{Cornelissen2017PRB,Oyanagi2020AIPAdv,Vlietstra2014PRB,Kikkawa2021NatCommun}; an a.c. charge current $I = \sqrt{2} I_{\rm rms} \sin (\omega t)$ at the frequency $\omega/2\pi$ of 13.423 Hz is applied to the injector Pt strip and a resultant second harmonic nonlocal voltage across the detector Pt strip is measured. Henceforth, we plot the nlSSE coefficient normalized by the applied current squared and the Pt length: $\tilde S \equiv V/(I_{\rm rms}^2 l_{V} )$ \cite{Cornelissen2015NatPhys,Cornelissen2016PRB-H-dep,Gomez-Perez2020PRB}. All the data are obtained in the linear regime \cite{Cornelissen2017PRB,Oyanagi2020AIPAdv}, where $V \propto I_{\rm rms}^2$. \par
%
%
\section{RESULTS AND DISCUSSION}
\subsection{Structural characterization} \label{sec:XRD-TEM}
Figure \ref{fig:TEM} shows the cross-sectional TEM images of the Pt/YIG/GGG, Pt/Bi$_{0.5}$Y$_{2.5}$IG/GGG, and  Pt/Bi$_{0.9}$Y$_{2.1}$IG/SGGG samples. 
The overall sample cross-section images [Figs. \ref{fig:TEM}(a), \ref{fig:TEM}(d), and \ref{fig:TEM}(g)] show high uniformity and flatness of the films.  
The high resolution images at the Bi$_x$Y$_{3-x}$IG/(S)GGG(111) interfaces  [Figs. \ref{fig:TEM}(c), \ref{fig:TEM}(f), and \ref{fig:TEM}(i)] reveal atomically smooth and epitaxial interfaces of our garnet films. Neither macroscopic defects nor misalignment in the lattice planes were observed in TEM images.
As shown in the insets of Figs. \ref{fig:TEM}(a), \ref{fig:TEM}(d), and \ref{fig:TEM}(g), the diffraction patterns of the Bi$_x$Y$_{3-x}$IG films coincide with those of the (S)GGG substrates, indicating that the Bi$_x$Y$_{3-x}$IG layers are grown as single-crystalline films with the [111] orientation in the out-of-plane direction. We also confirmed clear interfaces between the (polycrystalline) Pt-film and Bi$_x$Y$_{3-x}$IG-film contacts [see Figs. \ref{fig:TEM}(b), \ref{fig:TEM}(e), and \ref{fig:TEM}(h)], which allows us to investigate interfacial spin-current transport in these sample systems. \par
%
%
%
\begin{figure*}[htb]
\begin{center}
\includegraphics[width=14.5cm]{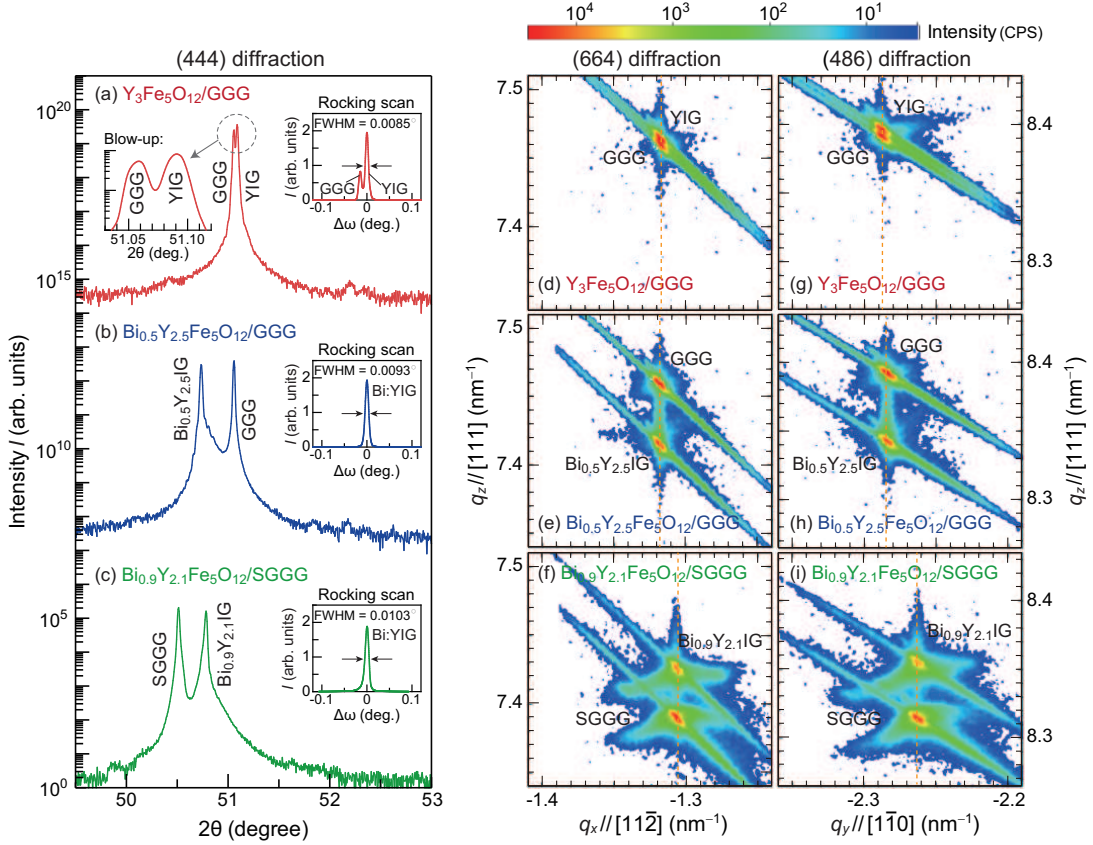}
\end{center}
\caption{
(a)-(c) $2\theta$-$\omega$ XRD around the (444) diffraction peaks of the (a) YIG($2.50~\mu \textrm{m}$)/GGG(111), (b) Bi$_{0.5}$Y$_{2.5}$IG($1.16~\mu \textrm{m}$)/GGG(111), and (c)  Bi$_{0.9}$Y$_{2.1}$IG($0.78~\mu \textrm{m}$)/SGGG(111) samples. 
The insets in (a), (b), and (c) show the rocking-scan results around the (444) diffraction peaks of the (a) YIG, (b) Bi$_{0.5}$Y$_{2.5}$IG, and (c) Bi$_{0.9}$Y$_{2.1}$IG films, from which the FWHM values were estimated to be (a) $0.0085^\circ$, (b) $0.0093^\circ$, and (c) $0.0103^\circ$, respectively.  
(d),(e),(f) [(g),(h),(i)] RSMs around the (664) [(486)] diffraction peaks of the (d) [(g)] YIG/GGG, (e) [(h)] Bi$_{0.5}$Y$_{2.5}$IG/GGG, and (f) [(i)]  Bi$_{0.9}$Y$_{2.1}$IG/SGGG samples. 
The RSMs show that the Bi$_x$Y$_{3-x}$IG films have a coherently strained pseudomorphic structure on the (S)GGG substrates (see the orange dashed lines, which show that the diffraction peak positions for the Bi$_x$Y$_{3-x}$IG film and substrate layers along the in-plane directions are the same for each sample \cite{Kubota2012APEX,Kubota2013JMMM}).  
}
\label{fig:XRD-RSM}
\end{figure*}
Figures \ref{fig:XRD-RSM}(a), \ref{fig:XRD-RSM}(b), and \ref{fig:XRD-RSM}(c) display the $2\theta$-$\omega$ XRD patterns of the Bi$_x$Y$_{3-x}$IG films around the (444) Bragg peaks from the GGG ($2\theta = 51.058^\circ$) or SGGG ($2\theta = 50.546^\circ$) substrates. 
For all the samples, we observed the (444) diffraction peaks from the Bi$_x$Y$_{3-x}$IG layer, confirming the out-of-plane orientation of Bi$_x$Y$_{3-x}$IG[111]/(S)GGG[111], consistent with the TEM results shown in Fig. \ref{fig:TEM}.  
The XRD rocking curves exhibit full width at half maximum (FWHM) values of $0.0085^\circ$, $0.0093^\circ$, and $0.0103^\circ$ for the YIG, Bi$_{0.5}$Y$_{2.5}$IG, and Bi$_{0.9}$Y$_{2.1}$IG, respectively, which are close to the values for the substrates ($0.0078^\circ$ for GGG and $0.0099^\circ$ for SGGG), ensuring high crystalline quality of our films. \par
To further characterize structural properties of the films, we performed reciprocal space mapping (RSM) \cite{Kubota2012APEX,Kubota2013JMMM,Avci2017NatMat}. 
Figures \ref{fig:XRD-RSM}(d), \ref{fig:XRD-RSM}(e), \ref{fig:XRD-RSM}(f) show the RSM data around the (664) diffraction for the YIG/GGG, Bi$_{0.5}$Y$_{2.5}$IG/GGG, and  Bi$_{0.9}$Y$_{2.1}$IG/SGGG samples, respectively. For all the samples, the asymmetric diffraction peaks lie on one vertical line, indicating that the films have a pseudomorphic structure with the in-plane lattice constant identical to that of the substrate along the $\langle 1 1 \overline{2} \rangle$ direction \cite{Kubota2012APEX,Kubota2013JMMM,Avci2017NatMat}.  
The pseudomorphic structure was confirmed also along the $\langle 1 \overline{1} 0 \rangle$ direction by the RSM data for the (486) diffraction [see Figs. \ref{fig:XRD-RSM}(g), \ref{fig:XRD-RSM}(h), and \ref{fig:XRD-RSM}(i)]. \par  
From the XRD and RSM results, the out-of-plane (in-plane) lattice constant was estimated to be
$12.376~{\rm \AA}$ ($12.383~{\rm \AA}$) for the YIG film on GGG, 
$12.456~{\rm \AA}$ ($12.383~{\rm \AA}$) for the Bi$_{0.5}$Y$_{2.5}$IG film on GGG, and 
$12.446~{\rm \AA}$ ($12.508~{\rm \AA}$) for the Bi$_{0.9}$Y$_{2.1}$IG film on SGGG. 
This result shows that the YIG film grown on GGG is nearly free from strain, while the Bi$_{0.5}$Y$_{2.5}$IG (Bi$_{0.9}$Y$_{2.1}$IG) film grown on GGG (SGGG) has compressive (tensile) epitaxial strain. 
In TABLE \ref{tab:comparison}, we show the in-plane biaxial strain $\epsilon_{||}$ and out-of-plane uniaxial strain $\epsilon_{\perp}$ values evaluated according to Ref. \onlinecite{Kubota2013JMMM}. 
The in-plane stress $\sigma_{||}$ values \cite{Kubota2013JMMM} are also calculated using the elastic coefficients $C_{11}\,(=\rho c_{\rm LA}^2$; $\rho$: mass density) and $C_{44}\,(=\rho c_{\rm TA}^2)$ and the relation $C_{11} - C_{12} = 2C_{44}$ \cite{Gurevich-Melkov_text,YIG_Paoletti_text}, which we list also in TABLE \ref{tab:comparison}. \par 
\begin{table*}[t]
\setlength{\belowcaptionskip}{0mm}
\caption{
Parameters for the (magneto)elastic properties of YIG, Bi$_{0.5}$Y$_{2.5}$IG, and Bi$_{0.9}$Y$_{2.1}$IG.
$\epsilon_{||(\perp)}$, $\sigma_{||}$, $c_{\rm TA (LA)}$, $\lambda_{111}$, and $B_{\perp }$ denote the in-plane biaxial (out-of-plane uniaxial) strain, in-plane stress, TA(LA)-phonon sound velocity, magnetostriction constant, and magnetoelastic coupling constant, respectively. 
$c_{\rm TA (LA)}$ values for the YIG film are adopted from Ref. \onlinecite{Gurevich-Melkov_text}. 
$\lambda_{111}$ values were evaluated from the relationship $\lambda_{111} = -2.819 \cdot (1+0.75x)  \times 10^{-6} $ for Bi$_{x}$Y$_{3-x}$IG films \cite{Soumah2018NatCommun}. 
$c_{\rm TA (LA)}$ values for Bi$_{0.5}$Y$_{2.5}$IG and Bi$_{0.9}$Y$_{2.1}$IG were measured by means of a time-resolved magneto-optical imaging method, followed by a Fourier transform process \cite{Hashimoto2017NatCommun,Hashimoto2018PRB}. 
$B_{\perp }$ values were calculated with the use of $c_{\rm TA}$ and $\lambda_{111}$ (see also the main text).
}   
\label{tab:comparison}
\begin{center}
\begin{tabular}{lllllllll} \hline \hline
Sample & ~~\,\,\,\,\,\, $\epsilon_{||}$       & ~~\,\,\,\,\,\, $\epsilon_{\perp}$       & ~~\,\,\,\,\,\, $\sigma_{||}$       & ~~\,\,\,\,\,\, $c_{\rm TA}$       &  ~~\,\,\,\,\,\, $c_{\rm LA}$              & ~~\,\,\,$\lambda_{111}$~~ & ~~\,$B_{\perp }/2\pi$~ \\
           & ~\,\,\,\,\,\, (\%)~~       & ~\,\,\,\,\,\, (\%)~~       & ~\,\,\, (GPa)~~        &~($10^{3}~\text{m}/\text{s}$)~~&~($10^{3}~\text{m}/\text{s}$)~~ &~~($10^{-6}$)~             &  ~~~($\text{THz}$)~  \\\hline 
Y$_{3}$Fe$_{5}$O$_{12}$                & ~~~~$0.032$  & ~~$-0.027$  & ~~~~$0.092$  & ~~~~~$3.84$  & ~~~~\,$7.21$ & ~\,$-2.82$ & ~~~\,\,$1.87$ \\
Bi$_{0.5}$Y$_{2.5}$Fe$_{5}$O$_{12}$ & ~~~$-0.31$  & ~~~~\,$0.28$  & ~~\,\,$-0.72$  & ~~~~~$3.34$  & ~~~~\,$6.42$ & ~\,$-3.88$ & ~~~\,\,$2.05$ \\ 
Bi$_{0.9}$Y$_{2.1}$Fe$_{5}$O$_{12}$ & ~~~\,\,\,$0.27$  & ~~\,$-0.24$  & ~\,\,\,\,\,\,\,$0.63$  & ~~~~~$3.30$  & ~~~~\,$6.28$ & ~\,$-4.72$ & ~~~\,\,$2.57$ \\ \hline \hline
\end{tabular}
\end{center}
\end{table*}
\subsection{Magnetic characterization} \label{sec:magnetization}
\begin{figure}[htb]
\begin{center}
\includegraphics[width=8.5cm]{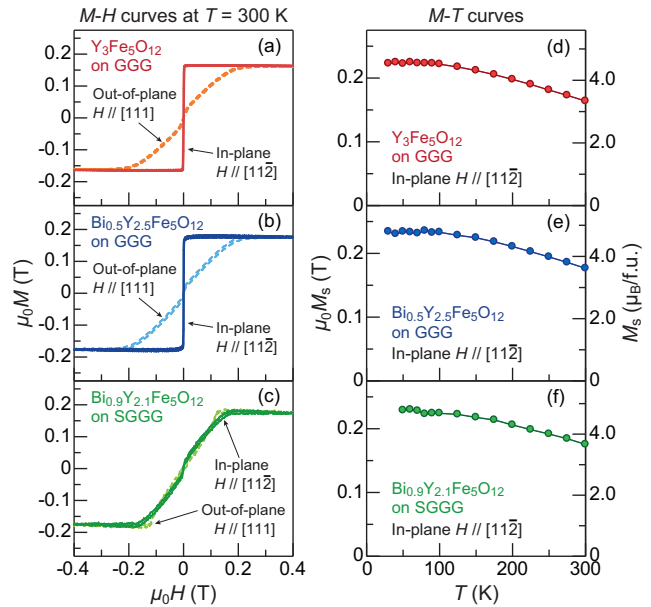}
\end{center}
\caption{(a)-(c) 
$H$ dependence of $M$ of the (a) YIG, (b) Bi$_{0.5}$Y$_{2.5}$IG, and (c) Bi$_{0.9}$Y$_{2.1}$IG films measured at $T=300~\textrm{K}$, where the magnetic field ${\bf H}$ was applied along the in-plane $[11\overline{2}]$ direction (solid lines) and along the out-of-plane $[111]$ direction (dashed lines). 
(d)-(f) $T$ dependence of the saturation magnetization $M_{\rm s}$ of the (d) YIG, (e) Bi$_{0.5}$Y$_{2.5}$IG, and (f) Bi$_{0.9}$Y$_{2.1}$IG films  
for ${\bf H}\;||\;[11\overline{2}]$.  
The $M$ values of these Bi$_x$Y$_{3-x}$IG films were extracted by
subtracting the contributions from the paramagnetic (S)GGG substrates.  Because of the large paramagnetic offset coming from the (S)GGG 
substrates at a low-$T$ range, the $M$ data were detectable at $T\geq30~\textrm{K}$ for the YIG and Bi$_{0.5}$Y$_{2.5}$IG films and at $T\geq50~\textrm{K}$ for the Bi$_{0.9}$Y$_{2.1}$IG film.
}
\label{fig:VSM}
\end{figure}

%
%
%
In Figs. \ref{fig:VSM}(a), \ref{fig:VSM}(b), and  \ref{fig:VSM}(c), we show the magnetic field $H$ dependence of the magnetization $M$ of the (a) YIG, (b) Bi$_{0.5}$Y$_{2.5}$IG, and (c) Bi$_{0.9}$Y$_{2.1}$IG films, respectively. 
Here, the $M$-$H$ curves are recorded at $300~\textrm{K}$ under the applied $H$ parallel to the in-plane $[11\overline{2}]$ (solid lines) and to the out-of-plane $[111]$ (dashed lines) axes. 
The pure YIG film (nearly free from strain) has an in-plane easy axis due to the shape anisotropy and negligibly-small magnetocrystalline anisotropy \cite{Kubota2013JMMM,SSE_Kikkawa2013PRL,Gallagher2016APL}; the saturation field $H_{\rm s}$ value for the in-plane $[11\overline{2}]$ field ($\mu_0 H_{\rm s}^{||} = 3~\textrm{mT}$) is much smaller than that for the out-of-plane $[111]$ field ($\mu_0 H_{\rm s}^{\perp} = 0.22~\textrm{T}$) [see Fig. \ref{fig:VSM}(a)].  
The in-plane saturation field $H_{\rm s}^{||}$ for the Bi$_{0.5}$Y$_{2.5}$IG (Bi$_{0.9}$Y$_{2.1}$IG) film is estimated to be $\mu_0 H_{\rm s}^{||} = 1~\textrm{mT}$ (0.17~\textrm{T}), which is much smaller (slightly higher) than that for the out-of-plane $\mu_0 H_{\rm s}^{\perp} = 0.23~\textrm{T}$ ($0.12~\textrm{T}$) [see Figs. \ref{fig:VSM}(b) and \ref{fig:VSM}(c)]. 
The different $H_{\rm s}^{||}$ v.s $H_{\rm s}^{\perp}$ features can be interpreted in terms of the stress-induced anisotropy;  
for a material with a negative magnetostriction constant $\lambda _{111} < 0$ such as Bi$_x$Y$_{3-x}$IG,  the magnetic easy axis tends to lie in (perpendicular to) the film plane for compressive (tensile) epitaxial strain \cite{Hansen1983PRB,Kubota2012APEX,Kubota2013JMMM,Tang2016PRB,Fu2017APL}.
Therefore, for the Bi$_{0.5}$Y$_{2.5}$IG films on GGG with the compressive epitaxial strain, both the shape anisotropy and stress-induced anisotropy make the magnetic easy axis within an in-plane direction ($H_{\rm s}^{||} \ll H_{\rm s}^{\perp}$).  
In contrast, for the Bi$_{0.9}$Y$_{2.1}$IG film on SGGG with the tensile epitaxial strain, the stress-induced out-of-plane anisotropy appears and prevails against the shape anisotropy, which renders the magnetic easy axis perpendicular to the film plane ($H_{\rm s}^{||} > H_{\rm s}^{\perp}$) \cite{Soumah2018NatCommun}. \par
The saturation magnetization $M_{\rm s}$ values were estimated to be $3.34~\mu_\textrm{B}$ ($0.164~\textrm{T}$), $3.62~\mu_\textrm{B}$ ($0.177~\textrm{T}$), and $3.66~\mu_\textrm{B}$ ($0.175~\textrm{T}$) at $300~\textrm{K}$, for the YIG, Bi$_{0.5}$Y$_{2.5}$IG, and Bi$_{0.9}$Y$_{2.1}$IG films, respectively, in units of the Bohr magneton (Tesla)  (see Fig. \ref{fig:VSM}).
The $M_{\rm s}$ value increases by the Bi substitution, which is consistent with the previous reports  \cite{Hansen1983PRB,Hansen1984ThinSolidFilms}.  
Figures \ref{fig:VSM}(d), \ref{fig:VSM}(e), and  \ref{fig:VSM}(f) show the $T$ dependence of $M_{\rm s}$ of the (d) YIG, (e) Bi$_{0.5}$Y$_{2.5}$IG, and (f) Bi$_{0.9}$Y$_{2.1}$IG films, respectively, for the $H$ direction parallel to the in-plane $[11\overline{2}]$ axis. For all the samples, the $M_{\rm s}$ value increases with decreasing $T$ and approaches to $\sim 5~\mu_{\rm B}$ at the lowest $T$, showing good agreement with the theoretical and previous experimental results \cite{YIG_Gilleo-Geller,Hansen1983PRB,Hansen1984ThinSolidFilms,Kikkawa2017PRB}.  
%
\subsection{Evaluation of sound velocity and magnetoelastic coupling} \label{sec:MEC}
\begin{figure}[htb]
\begin{center}
\includegraphics[width=8.5cm]{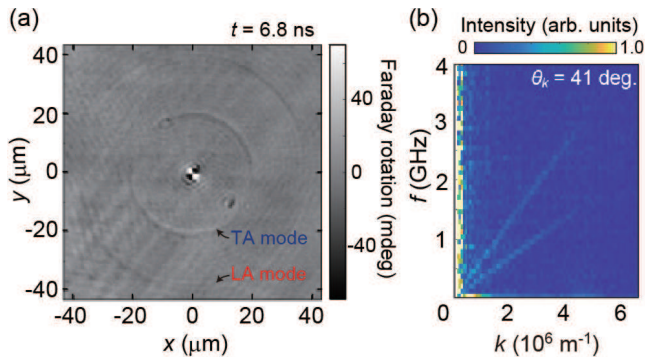}
\end{center}
\caption{(a) A time-resolved magneto-optical image for the Bi$_{0.9}$Y$_{2.1}$IG sample obtained at the time delay of 6.8 ns. The sample is magnetized slightly along the $x$ direction by the in-plane  external field of $\mu_0 H = 2.8~\textrm{mT}$, while the out-of-plane component of the magnetization distribution is measured through the Faraday effect of the probe pulse \cite{Hashimoto2017NatCommun,Hashimoto2018PRB}.
(b) The phonon dispersion relations for the Bi$_{0.9}$Y$_{2.1}$IG sample reconstructed by Fourier transforming the propagating waveform shown in (a) for $\theta_k = 41^\circ$, where $\theta_k$ denotes the angle between the wavevector ${\bf k}$ and ${\bf H} ~ ||+{\hat {\bf x}}$. 
}
\label{fig:SWaT}
\end{figure}
 
%
To determine the sound velocities of the Bi$_x$Y$_{3-x}$IG films, we employed a pump-and-probe magneto-optical imaging method combined with a Fourier transform (FT) process (see Refs. \onlinecite{Hashimoto2017NatCommun} and \onlinecite{Hashimoto2018PRB} for the details of the optical system in this experiment). 
When a pumped laser pulse with a spot radius of $\sim 2.5~\mu\textrm{m}$ is irradiated to the sample, elastic waves are excited and propagate radially from the excitation point. Via the magnetoelastic interaction, the elastic waves tilt the magnetic moments from an in-plane to out-of-plane direction that can be detected as a change of the Faraday rotation angle of the probe pulse \cite{Hashimoto2017NatCommun,Hashimoto2018PRB}. 
Figure \ref{fig:SWaT}(a) shows a snapshot of the temporal evolution of out-of-plane magnetization distribution of the Bi$_{0.9}$Y$_{2.1}$IG sample obtained at the time delay of $t = 6.8$ ns between the pump and
probe pulses.   
Two ring structures are visible; the inner (outer) ring is attributed to the coupled TA (LA) and spin waves having a smaller (larger) velocity.   
By Fourier transforming the observed propagating waveform with respect to the time and spatial coordinates, we obtain coupled phonon and spin waves dispersion relations.
As shown in Fig. \ref{fig:SWaT}(b), two $k$-linear dispersions with different slopes are discerned; the dispersion showing the lower (larger) slope is assigned to the TA (LA) phonon branch, from which we can estimate the sound velocities of each phonon mode. Measurements were performed also for the Bi$_{0.5}$Y$_{2.5}$IG sample. 
We note that the crossings between phonon and spin wave branches are inaccessible, since their positions in the momentum direction are large and therefore out of range in the present experiment. \par 
Significant decreases in the sound velocities for the Bi:YIG films are indeed confirmed through the above measurements and analysis.  
As shown in Table \ref{tab:comparison}, the TA- and LA-phonon sound velocities of the Bi$_{0.5}$Y$_{2.5}$IG films were determined to be $c_{\rm TA} = 3.34 \times 10^{3}~\textrm{m/s}$ and $c_{\rm LA} = 6.42 \times 10^{3}~\textrm{m/s}$, giving values smaller than those of YIG.
For the Bi$_{0.9}$Y$_{2.1}$IG film, $c_{\rm TA} = 3.30 \times 10^{3}~\textrm{m/s}$ and $c_{\rm LA} = 6.28 \times 10^{3}~\textrm{m/s}$, showing the smallest values among our samples. \par
Using the magnetostriction constant $\lambda_{111}$ values for $\mu$m-thick Bi$_x$Y$_{3-x}$IG films [$\lambda_{111} = -2.819 \cdot (1+0.75x)  \times 10^{-6} $] shown in Ref. \onlinecite{Soumah2018NatCommun} and the sound velocities $c_{\rm TA} \, (=\sqrt{C_{44}/\rho})$ in Table \ref{tab:comparison}, the magnetoelastic coupling constant $B_{\perp}/2\pi \, [=-3\,a_{\rm Bi:YIG}^3 \lambda_{111} C_{44}/(2\pi\hbar) \,]$ ($a_{\rm Bi:YIG}$: lattice constant) \cite{Kittel1949RevModPhys,Gurevich-Melkov_text,YIG_Paoletti_text} was estimated to be $1.87$, $2.05$, and $2.57~\textrm{THz}$ for the YIG, Bi$_{0.5}$Y$_{2.5}$IG, and Bi$_{0.9}$Y$_{2.1}$IG films, respectively.
The $B_{\perp}$ value monotonically increases by increasing the Bi amount $x$, which may be attributed to the enhanced spin-orbit coupling induced by the substitution of Bi$^{3+}$ ions \cite{Kumar2019JPCM}. \par  
%
%
\subsection{Magnon-polaron features in LSSE} \label{sec:LSSE}
\begin{figure}[htb]
\begin{center}
\includegraphics[width=8.5cm]{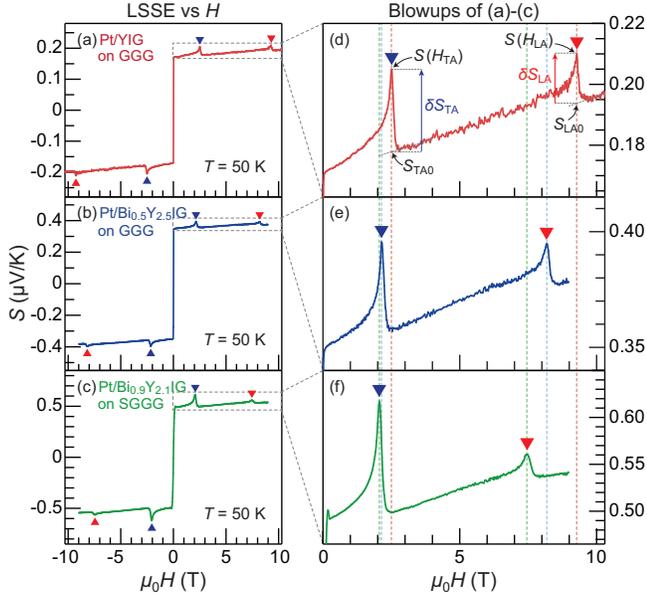}
\end{center}
\caption{(a)-(c) 
$H$ dependence of the LSSE coefficient $S$ of the (a) Pt/YIG, (b) Pt/Bi$_{0.5}$Y$_{2.5}$IG, and (c) Pt/Bi$_{0.9}$Y$_{2.1}$IG samples at $T=50~\textrm{K}$. 
(d)-(f) Corresponding magnified views of $S\left( H\right)$ around the anomaly fields. 
The $S$ peaks at $H_{\mathrm{TA}}$ and $H_{\mathrm{LA}}$ are marked by blue and red filled triangles, respectively. 
The light-red, light-blue, light-green dashed lines represent the peak positions for the Pt/YIG, Pt/Bi$_{0.5}$Y$_{2.5}$IG, and Pt/Bi$_{0.9}$Y$_{2.1}$IG samples, respectively.  
$S(H_i)$ and $S_{i0}$ ($i = $ TA or LA) shown in (d) represent the intensity of $S$ and extrapolated background $S$ at the peak position $H_i$, respectively, from which the anomaly intensity is evaluated as $\delta S_i \equiv S(H_i) - S_{i0}$.
The intensity of magnon-polaron anomalies relative to the background $S_{i0}$ is defined as $\delta S_i/S_{i0} \equiv [S(H_i) - S_{i0}]/S_{i0}$.  
}
\label{fig:LSSE-results-at-50K}
\end{figure}

%
%
%
Now, we present the experimental results on the LSSE in the Pt/Bi$_x$Y$_{3-x}$IG samples. 
Figure \ref{fig:LSSE-results-at-50K}(a) shows the $H$ dependence of the LSSE coefficient $S$ of the Pt/YIG sample measured at $T=50~\textrm{K}$. 
In this conventional system, we observed a clear SSE signal as well as two magnon-polaron-induced spikes on top of the almost flat $S$ background signal [see the blue and red filled triangles in Fig. \ref{fig:LSSE-results-at-50K}(a) and magnified view shown in Fig. \ref{fig:LSSE-results-at-50K}(d)]. 
The peak at the lower (higher) $H$ appears at $\mu _0 H_{\rm TA} = 2.51~\textrm{T}$ ($\mu _0 H_{\rm LA} = 9.27~\textrm{T}$), consistent with Ref. \onlinecite{Kikkawa2016PRL}. 
The $H_{\rm TA}$ and $H_{\rm LA}$ values coincide with the $H$ intensities at which the magnon dispersion curve touches the TA- and LA-phonon dispersion curves of YIG, respectively, being the conditions that the magnon and phonon modes can be coupled
over the largest volume in momentum space, so the magnon-polaron effects are maximized \cite{Kikkawa2016PRL,Flebus2017PRB}. \par 
Next, let us focus on the effect of Bi substitution on the LSSE and magnon-polaron features. 
Figures \ref{fig:LSSE-results-at-50K}(b) and \ref{fig:LSSE-results-at-50K}(c) show the $H$ dependence of $S$ of the Pt/Bi$_{0.5}$Y$_{2.5}$IG and Pt/Bi$_{0.9}$Y$_{2.1}$IG samples, respectively. We observed SSE signals in these samples, of which the amplitudes are comparable to that for the Pt/YIG shown in Fig. \ref{fig:LSSE-results-at-50K}(a). 
The peak features show up also in these Pt/Bi:YIG systems as marked by the blue and red filled triangles in Figs. \ref{fig:LSSE-results-at-50K}(b) and \ref{fig:LSSE-results-at-50K}(c).  
Interestingly, the peak positions shift toward lower $H$ values by increasing the amount of Bi substitution.
The peak positions at the lower and higher $H$ for the Pt/Bi$_{0.5}$Y$_{2.5}$IG sample are estimated to be $\mu _0 H_{\rm TA} = 2.15~\textrm{T}$ and $\mu _0 H_{\rm LA} = 8.18~\textrm{T}$, respectively. 
For the Pt/Bi$_{0.9}$Y$_{2.1}$IG, the peak fields are further decreased: $\mu _0 H_{\rm TA} = 2.06~\textrm{T}$ and $\mu _0 H_{\rm LA} = 7.47~\textrm{T}$. \par 
The observed peak shifts are explained in terms of the reduction of the sound velocities and resultant touching fields of Bi:YIG films.  
The touching field can be evaluated quantitatively by solving the combined equation for the magnon ($\omega_{\rm mag}$) and phonon ($\omega_{\rm ph}$) dispersion relations and their group velocities:
\begin{equation}
\omega_{\rm mag} = \omega_{\rm ph},  \;\; {\partial}\omega_{\rm mag}/{\partial k} =  {\partial}\omega_{\rm ph}/{\partial k} ,  \label{equ:coupled-eq}
\end{equation}
where the magnon dispersion relation, disregarding the dipolar interaction ($M _{\rm s} \ll H _{\rm TA, LA}$), reads $\omega_{\rm mag} = D_{\rm ex} k^2 + \gamma \mu _0 H$, while the phonon dispersions are given by $\omega_{\rm ph} = c_{\rm TA, LA} k$. Here, $D_{\rm ex}$, $k$, and $\gamma$ represent the exchange stiffness, wavenumber, and gyromagnetic ratio, respectively. 
Calculation of Eq. (\ref{equ:coupled-eq}) leads to
\begin{equation}
\mu _0 H _{\rm TA(LA)} = \frac{c_{\rm TA (LA)}^2}{4D_{\rm ex}\gamma}, \label{equ:touching-field}
\end{equation}
which shows that the touching field $H _{\rm TA(LA)}$ scales quadratically with the sound velocity $c_{\rm TA (LA)}$, and so does the anomaly field in the SSE.  
In fact, the observed peak fields $H_{\rm TA,LA}$ are well reproduced by the sound velocities $c_{\rm TA (LA)}$ shown in Table \ref{tab:comparison} and reasonable $D_{\rm ex}$ values of $8.0 \times 10^{-6}$, $7.1 \times 10^{-6}$, and $7.3 \times 10^{-6}~\textrm{m}^2/\textrm{s}$ for the YIG,  Bi$_{0.5}$Y$_{2.5}$IG, and Bi$_{0.9}$Y$_{2.1}$IG, respectively.  
The frequency value for the touching point $\omega_{\rm MTA}/2\pi$ ($\omega_{\rm MLA}/2\pi$) between the magnon and TA(LA)-phonon branches at $H_{\rm TA}$ ($H_{\rm LA}$), depicted as green filled circles in Fig. \ref{fig:BIYIG-dispersions-schematics}(b), can be evaluated as 
$\omega_{\rm MTA}/2\pi = 0.15~\textrm{THz}$ ($\omega_{\rm MLA}/2\pi = 0.52~\textrm{THz}$) for the YIG, 
$0.13~\textrm{THz}$ ($0.47~\textrm{THz}$) for the Bi$_{0.5}$Y$_{2.5}$IG, and $0.12~\textrm{THz}$ ($0.43~\textrm{THz}$) for the Bi$_{0.9}$Y$_{2.1}$IG. \par  
%
%
%
%
%
\begin{figure*}[tbh]
\begin{center}
\includegraphics[width=16cm]{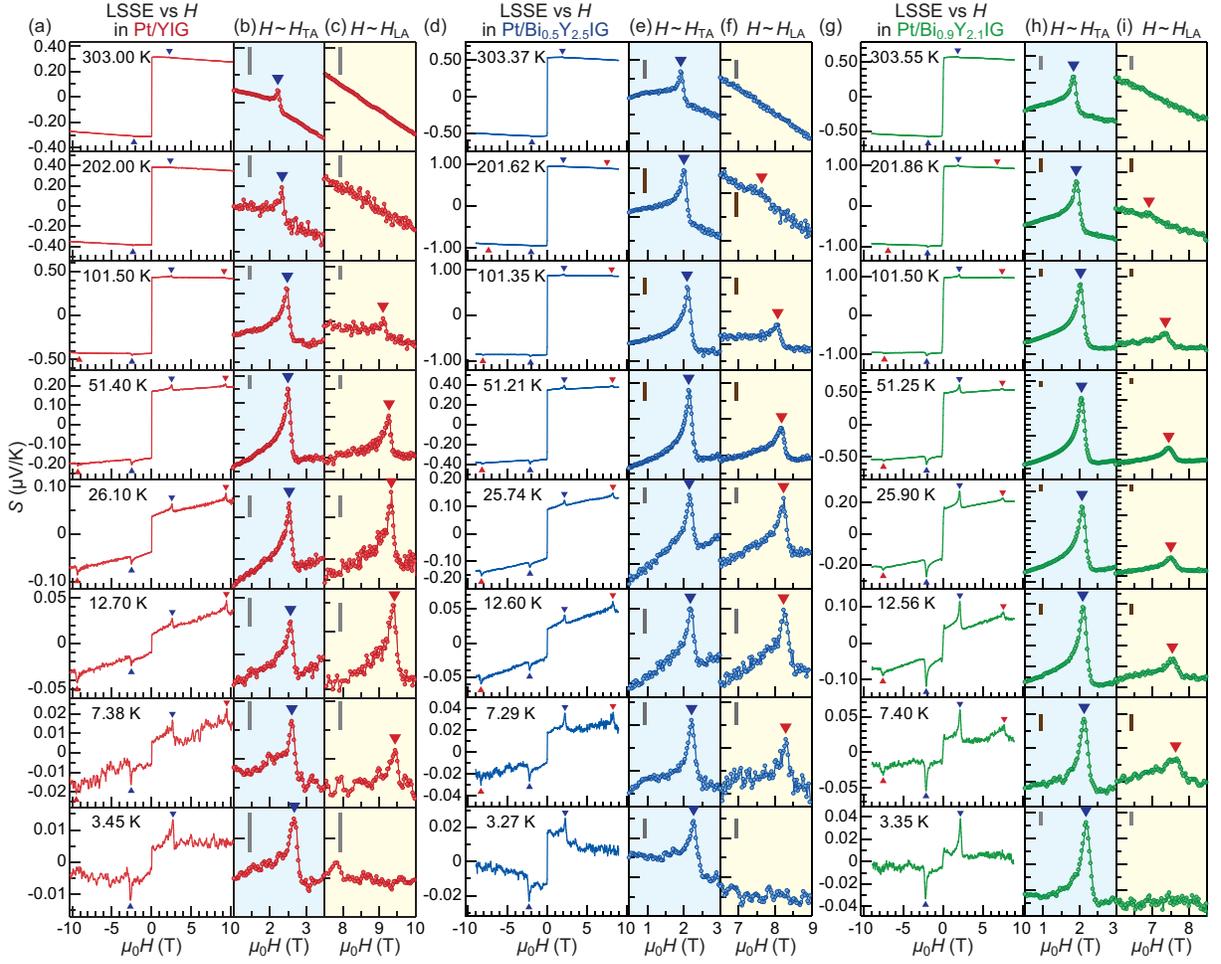}
\end{center}
\caption{
(a),(d),(g) $S\left(H\right) $ of the  (a) Pt/YIG, (d) Pt/Bi$_{0.5}$Y$_{2.5}$IG, and (g) Pt/Bi$_{0.9}$Y$_{2.1}$IG samples for various values of the sample temperature $T_{\mathrm{avg}} =(T_{\rm H} + T_{\rm L})/2$.  
(b),(e),(h) [(c),(f),(i)] Blowups of $S$ versus $H$ around $H_{\mathrm{TA}}$ ($H_{\mathrm{LA}}$) for the (b) [(c)] Pt/YIG, (e) [(f)] Pt/Bi$_{0.5}$Y$_{2.5}$IG, and (h) [(i)] Pt/Bi$_{0.9}$Y$_{2.1}$IG samples.
The gray and brown scale bars represent $0.005~\mu \textrm{V/K}$ and $0.010~\mu \textrm{V/K}$, respectively.
The $S$ peaks at $H_{\mathrm{TA}}$ and $H_{\mathrm{LA}}$ are marked by blue and red filled triangles, respectively.
The global $S$ signal below $\sim 30~\textrm{K}$ increases with increasing $H$, which may be attributed to the additional spin-current generation by the paramagnetic (S)GGG substrate \cite{Kikkawa2016PRL,SSE_Wu2015PRL,Chen2019AIPAdv,Oyanagi2019NatCommun,Oyanagi2021PRB,Oyanagi2022}. 
}
\label{LSSE-anomalies-in-Pt-YIG-BiYIG-films-T-dep-scale-bars-only}
\end{figure*}
\begin{figure*}[tbh]
\begin{center}
\includegraphics{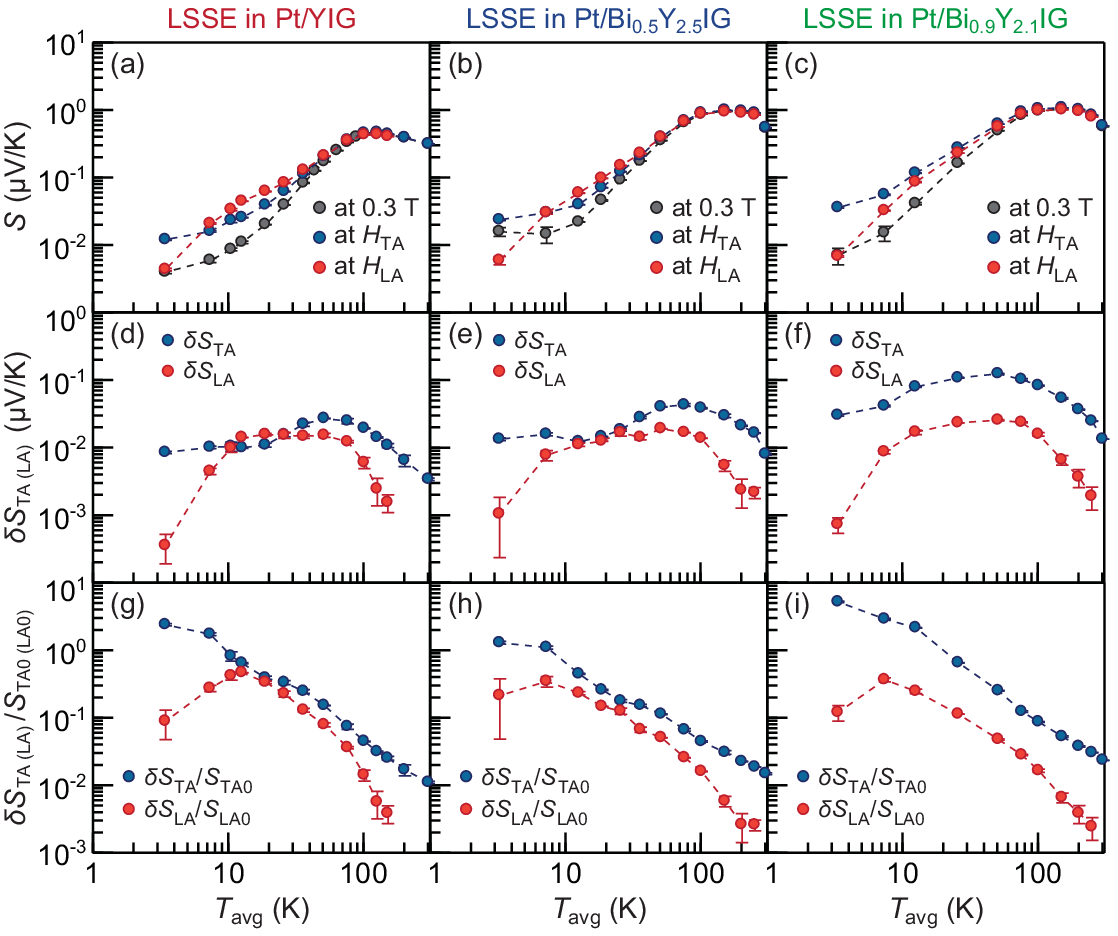}
\end{center}
\caption{
(a),(b),(c) 
$T_{\mathrm{avg}}$ dependence of $S$ at $\mu_{0} H=0.3~\text{T}$, $\mu_{0} H_{\mathrm{TA}}$, and $\mu_{0} H_{\mathrm{LA}}$ for the (a) Pt/YIG, (b) Pt/Bi$_{0.5}$Y$_{2.5}$IG, and (c) Pt/Bi$_{0.9}$Y$_{2.1}$IG samples. 
(d),(e),(f) [(g),(h),(i)]
$T_{\mathrm{avg}}$ dependence of the intensities of magnon-polaron peaks $\delta S_{\mathrm{TA}}$ and $\delta S_{\mathrm{LA}}$ [$\delta S_{\mathrm{TA}}/S_{\mathrm{TA0}}$ and $\delta S_{\mathrm{LA}}/S_{\mathrm{LA0}}$] for the (d) [(g)] Pt/YIG, (e) [(h)] Pt/Bi$_{0.5}$Y$_{2.5}$IG, and (f) [(i)] Pt/Bi$_{0.9}$Y$_{2.1}$IG samples.  
The definitions of $\delta S_{\mathrm{TA}}$, $\delta S_{\mathrm{LA}}$, $\delta S_{\mathrm{TA}}/S_{\mathrm{TA0}}$, and $\delta S_{\mathrm{LA}}/S_{\mathrm{LA0}}$ are shown in the caption of Fig. \ref{fig:LSSE-results-at-50K}.
}
\label{fig:LSSE-anomalies-in-Pt-YIG-BiYIG-films-T-dep-summary}
\end{figure*}

We carried out systematic measurements of the temperature dependence of the LSSE in these Pt/Bi$_x$Y$_{3-x}$IG samples. 
Figures \ref{LSSE-anomalies-in-Pt-YIG-BiYIG-films-T-dep-scale-bars-only}(a),  \ref{LSSE-anomalies-in-Pt-YIG-BiYIG-films-T-dep-scale-bars-only}(d), and \ref{LSSE-anomalies-in-Pt-YIG-BiYIG-films-T-dep-scale-bars-only}(g) show the $S$-$H$ curves for the various average sample temperature $T_{\mathrm{avg}}$ (from $300$ to $3~\textrm{K}$) for the Pt/YIG, Pt/Bi$_{0.5}$Y$_{2.5}$IG, and Pt/Bi$_{0.9}$Y$_{2.1}$IG samples, respectively. 
In all the samples, the overall $T$ dependences of $S$ agree with those for the Pt/YIG-film samples reported in Refs. \onlinecite{Kikkawa2015PRB} and \onlinecite{Jin2015PRB};  
when the sample temperature is decreased from $300~\textrm{K}$, the magnitude of $S$ increases and shows a broad peak at around $T = 150~\textrm{K}$ [see Figs. \ref{fig:LSSE-anomalies-in-Pt-YIG-BiYIG-films-T-dep-summary}(a)-\ref{fig:LSSE-anomalies-in-Pt-YIG-BiYIG-films-T-dep-summary}(c)]. On decreasing $T$ further, the $S$ signal begins to decrease and eventually goes to zero. \par
%
We now discuss the $T$ dependence of the magnon-polaron features in the LSSE. 
First, let us focus on the results at a low-$T$ range  below $50~\textrm{K}$,
at which characteristic $T$ response is expected to occur due to the competition between the thermal energy $k_{\rm B}T$ and the magnon-polaron's touching energy $\hbar \omega_{\mathrm{MTA,MLA}}$, which governs the $T$ dependence of thermal occupation of magnon-polaron modes \cite{Kikkawa2016PRL}.   
Figures \ref{fig:LSSE-anomalies-in-Pt-YIG-BiYIG-films-T-dep-summary}(d), \ref{fig:LSSE-anomalies-in-Pt-YIG-BiYIG-films-T-dep-summary}(e), and \ref{fig:LSSE-anomalies-in-Pt-YIG-BiYIG-films-T-dep-summary}(f) represent the $T$ dependence of the magnon-polaron peak amplitude $\delta S_{\rm TA,LA} \equiv S(H_{\rm TA,LA}) - S_{\rm TA0, LA0}$ for the Pt/YIG, Pt/Bi$_{0.5}$Y$_{2.5}$IG, and Pt/Bi$_{0.9}$Y$_{2.1}$IG samples, respectively, where $S_{\rm TA0(LA0)}$ is the extrapolated background $S$ at the peak position $H_{\rm TA(LA)}$ [see also the definition shown in Fig. \ref{fig:LSSE-results-at-50K}(d) and its caption]. 
The peak intensities $\delta S_{\rm TA,LA}$ at $H = H_{\mathrm{TA}}$ and $H_{\mathrm{LA}}$ exhibit different $T$ dependences.  
We further introduce and hereafter discuss the quantity $\delta S_{\mathrm{TA(LA)}}/S_{\mathrm{TA0(LA0)}}$, where $\delta S_{\rm TA(LA)}$ is normalized by the background $S_{\rm TA0(LA0)}$, which allows us to evaluate the $T$-dependence of the magnon-polaron SSE coefficient relative to the background magnonic one at  $H = H_{\mathrm{TA(LA)}}$ and also to exclude common $T$-dependent factors in the magnon-polaron and magnonic SSE voltages such as the interfacial spin conductance, electric resistance, and spin diffusion length in the Pt film \cite{Cornelissen2016PRB-chemical-potential,Guo2016PRX,Cornelissen2017PRB}. 
As shown in Figs. \ref{fig:LSSE-anomalies-in-Pt-YIG-BiYIG-films-T-dep-summary}(g)-\ref{fig:LSSE-anomalies-in-Pt-YIG-BiYIG-films-T-dep-summary}(i),
$\delta S_{\mathrm{TA}}/S_{\mathrm{TA0}}$ monotonically increases with decreasing $T$ and takes a maximum value at the lowest $T$ for all the samples (see also the $S$-$H$ curves shown in Fig. \ref{LSSE-anomalies-in-Pt-YIG-BiYIG-films-T-dep-scale-bars-only}).
In contrast,  $\delta S_{\mathrm{LA}}/S_{\mathrm{LA0}}$, is gradually suppressed below $T^{\ast}  \sim  12~\text{K}$ ($\sim 7~\text{K}$) for the Pt/YIG sample (Pt/Bi$_{0.5}$Y$_{2.5}$IG and Pt/Bi$_{0.9}$Y$_{2.1}$IG samples) [see Figs. \ref{LSSE-anomalies-in-Pt-YIG-BiYIG-films-T-dep-scale-bars-only} and \ref{fig:LSSE-anomalies-in-Pt-YIG-BiYIG-films-T-dep-summary}(g)-\ref{fig:LSSE-anomalies-in-Pt-YIG-BiYIG-films-T-dep-summary}(i)]. 
At the lowest $T$, the $S$ peak at $H_{\mathrm{LA}}$ is so small and almost indistinguishable from the background [see Figs. \ref{LSSE-anomalies-in-Pt-YIG-BiYIG-films-T-dep-scale-bars-only}(c), \ref{LSSE-anomalies-in-Pt-YIG-BiYIG-films-T-dep-scale-bars-only}(f), and \ref{LSSE-anomalies-in-Pt-YIG-BiYIG-films-T-dep-scale-bars-only}(i)]. 
The $T$-dependent feature can be interpreted in terms of the difference in the frequency values 
of the branch touching point $\omega_{\mathrm{MTA}}$ for $H=H_{\mathrm{TA}}$ and $\omega_{\mathrm{MLA}}$ for $H=H_{\mathrm{LA}}$ [see Fig. \ref{fig:BIYIG-dispersions-schematics}(b)] \cite{Kikkawa2016PRL,Flebus2017PRB}. 
Here, in the unit of temperature, the $\omega_{\mathrm{MLA}}$ values approximately correspond to $T_{\rm MLA} (= \hbar \omega_{\rm MLA}/k_{\rm B}) = 25~\textrm{K}$, $22~\textrm{K}$, and $21~\textrm{K}$ for the YIG, Bi$_{0.5}$Y$_{2.5}$IG, and Bi$_{0.9}$Y$_{2.1}$IG, respectively, and they are more than 3 times greater than those of the $\omega_{\mathrm{MTA}}$ values ($7~\textrm{K}$, $6~\textrm{K}$, and $6~\textrm{K}$, respectively).  
Therefore, for $T < T_{\rm MLA}$, the excitation of magnon polarons at the touching frequency $\omega \sim \omega_{\mathrm{MLA}}$ is rapidly suppressed, which leads to the disappearance of the $S$ peak at  $H_{\mathrm{LA}}$ at the lowest $T$.  
Furthermore, the reduction of $\omega_{\mathrm{MLA}}$ values for Bi$_{0.5}$Y$_{2.5}$IG and Bi$_{0.9}$Y$_{2.1}$IG, by a value of $\sim 5~\textrm{K}$, compared to that for the YIG may be responsible for the difference of the onset $T^{\ast}$ values between the Pt/Bi:YIG ($T^{\ast} \sim 7~\textrm{K}$) and Pt/YIG ($T^{\ast} \sim 12~\textrm{K}$),  at which $\delta S_{\mathrm{LA}}/S_{\mathrm{LA0}}$ starts to be suppressed with decreasing $T$. \par
Next, we move on to the results at a higher-$T$ range ($50~\textrm{K} \lesssim T_{\rm avg} \lesssim 300~\textrm{K}$). 
In this $T$ range, we are also able to resolve small but finite peak structures at $H = H_{\mathrm{TA}}$ for all the samples, while, at $H_{\mathrm{LA}}$, the peak structures were detectable below $\sim 150~\textrm{K}$ ($\sim 250~\textrm{K}$) for the Pt/YIG (Pt/Bi:YIG) sample (see Figs. \ref{LSSE-anomalies-in-Pt-YIG-BiYIG-films-T-dep-scale-bars-only} and \ref{fig:LSSE-anomalies-in-Pt-YIG-BiYIG-films-T-dep-summary}). 
As shown in Figs. \ref{fig:LSSE-anomalies-in-Pt-YIG-BiYIG-films-T-dep-summary}(g)-\ref{fig:LSSE-anomalies-in-Pt-YIG-BiYIG-films-T-dep-summary}(i), the peak amplitudes at $H_{\mathrm{TA}}$ and $H_{\mathrm{LA}}$ relative to the background SSE signals, $\delta S_{\mathrm{TA}}/S_{\mathrm{TA0}}$ and $\delta S_{\mathrm{LA}}/S_{\mathrm{LA0}}$, monotonically decrease with increasing $T$, which may be due to the appearance of large contributions of thermally populated (uncoupled) magnons with the energy of $k_{\rm B} T > \hbar \omega_{\rm MTA, MLA}$ to the background SSE signal. 
Furthermore, the observed structures with only peak shapes (no dip shapes) suggest that the scattering rate of magnons $\tau^{-1}_{\rm mag}$ with the frequencies relevant to the hybridization (i.e., $\omega \sim \omega_{\rm MTA, MLA}$) are always larger than that of phonons $\tau^{-1}_{\rm ph}$ at all the $T$ range: $\tau^{-1}_{\rm mag} > \tau^{-1}_{\rm ph}$ \cite{Kikkawa2016PRL,Flebus2017PRB}. 
We also found that, in this $T$ range, as the temperature is increased, the $\delta S_{\mathrm{LA}}/S_{\mathrm{LA0}}$ value decreases much faster than the $\delta S_{\mathrm{TA}}/S_{\mathrm{TA0}}$ value for all the samples [see Figs. \ref{fig:LSSE-anomalies-in-Pt-YIG-BiYIG-films-T-dep-summary}(g)-\ref{fig:LSSE-anomalies-in-Pt-YIG-BiYIG-films-T-dep-summary}(i)]. 
We note that with increasing $T$, dominant scattering sources for magnons and phonons may change from disorders (impurities) to inelastic magnon-magnon and phonon-phonon scatterings due to their increased thermal population \cite{Rezende2014PRB,Cornelissen2016PRB-chemical-potential,Boona2014PRB,Shi2021PRL_YIG-bulk,Schmidt2018PRB}.
A detailed knowledge on the $T$ and $k$ dependences of such scattering events for the Bi$_x$Y$_{3-x}$IG films will thus be important to fully understand the observed behavior. \par  
Finally, we compare the amplitudes of the peak structures in the LSSE in the Pt/Bi$_x$Y$_{3-x}$IG samples.  
The magnon-polaron contribution $\delta S_{\mathrm{TA}}/S_{\mathrm{TA0}}$ at $H=H_{\rm TA}$ for the Pt/YIG, Pt/Bi$_{0.5}$Y$_{2.5}$IG, and Pt/Bi$_{0.9}$Y$_{2.1}$IG samples at the lowest $T \sim 3~\textrm{K}$ ($T=50~\textrm{K}$) was evaluated as 2.4, 1.3, and 5.2 (0.15, 0.11, and 0.25), respectively.
This nonmonotonic dependence with respect to the Bi amount is not simply explained by the magnetoelastic coupling constant $B_{\perp}$, since it monotonically increases by increasing the Bi substitution (see TABLE \ref{tab:comparison}). Although larger $B_{\perp}$ values should be beneficial for the magnon-polaron formation and resultant SSE anomalies, our results suggest that the scattering ratio parameterized by $\eta = \tau^{-1}_{\rm mag}/\tau^{-1}_{\rm ph}$ may play a rather important role for the comparison of the magnon-polaron peak intensities between each sample. Furthermore, in the LSSE results, we found that the magnon-polaron contribution at $H=H_{\rm TA}$ is greater than that at $H_{\rm LA}$, i.e., $\delta S_{\mathrm{TA}}/S_{\mathrm{TA0}}>\delta S_{\mathrm{LA}}/S_{\mathrm{LA0}}$ for all the $T$ range and all the samples [see Figs. \ref{fig:LSSE-anomalies-in-Pt-YIG-BiYIG-films-T-dep-summary}(g)-\ref{fig:LSSE-anomalies-in-Pt-YIG-BiYIG-films-T-dep-summary}(i)]. This tendency is compared with that for the nlSSE results in the next section.  
%
%
\subsection{Magnon-polaron features in nlSSE} \label{sec:nlSSE}
%
\begin{figure}[htb]
\begin{center}
\includegraphics[width=8.5cm]{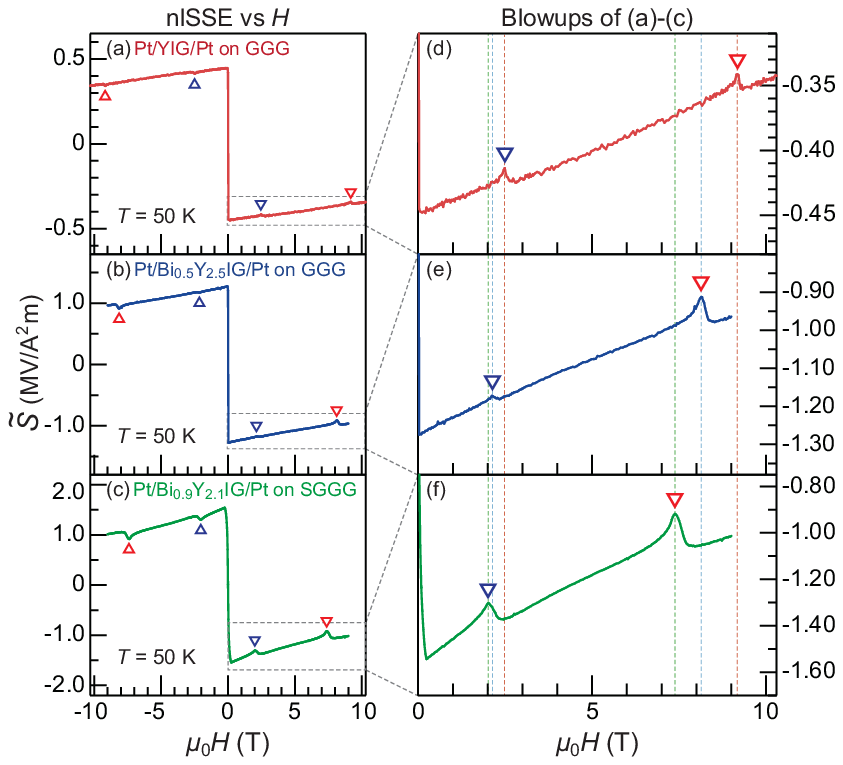}
\end{center}
\caption{(a)-(c) 
$H$ dependence of the nlSSE coefficient ${\tilde S}$ of the (a) Pt/YIG/Pt, (b) Pt/Bi$_{0.5}$Y$_{2.5}$IG/Pt, and (c) Pt/Bi$_{0.9}$Y$_{2.1}$IG/Pt samples at $T=50~\textrm{K}$. 
(d)-(f) Corresponding magnified views of ${\tilde S}\left( H\right)$ around the anomaly fields.  
The ${\tilde S}$ dips at $H_{\mathrm{TA}}$ and $H_{\mathrm{LA}}$ are marked by blue and red unfilled triangles, respectively. 
The light-red, light-blue, light-green dashed lines represent the peak positions for the Pt/YIG, Pt/Bi$_{0.5}$Y$_{2.5}$IG, and Pt/Bi$_{0.9}$Y$_{2.1}$IG samples, respectively. 
The intensity of magnon-polaron anomalies is evaluated as $\delta {\tilde S}_i \equiv {\tilde S}(H_i) - {\tilde S}_{i0}$  ($i = $ TA or LA), where ${\tilde S}(H_i)$ and ${\tilde S}_{i0}$ represent the intensity of ${\tilde S}$ and extrapolated background ${\tilde S}$ at the peak position $H_i$, respectively, similar to the case for the LSSE [see also the caption of Fig. \ref{fig:LSSE-results-at-50K}(d)]. 
The intensity of the anomaly relative to the background ${\tilde S}_{i0}$ is defined as $\delta {\tilde S}_i/{\tilde S}_{i0} \equiv [{\tilde S}(H_i) - {\tilde S}_{i0}]/{\tilde S}_{i0}$. 
}
\label{fig:nlSSE-results-at-50K}
\end{figure}
We now show the results on the nlSSE in the Pt/Bi$_x$Y$_{3-x}$IG/Pt systems. 
Figures \ref{fig:nlSSE-results-at-50K}(a), \ref{fig:nlSSE-results-at-50K}(b), and \ref{fig:nlSSE-results-at-50K}(c) show the $H$ dependence of the nlSSE coefficient ${\tilde S}$ of the Pt/YIG/Pt, Pt/Bi$_{0.5}$Y$_{2.5}$IG/Pt, and Pt/Bi$_{0.9}$Y$_{2.1}$IG/Pt samples measured at $T=50~\textrm{K}$, respectively. 
The overall signal appears with a sign opposite to that for the LSSE, indicating that thermally-excited magnons accumulate beneath the detector Pt strip. The observed sign is indeed a characteristic in nlSSEs for $\mu$m-thick garnet films with a long injector-detector separation distance $d$ ($d = 8~\mu\textrm{m}$ in the present case) \cite{Cornelissen2017PRB,Shan2016PRB,Shan2017PRB}.  
As marked by the blue and red unfilled triangles in the nlSSE signals in Fig. \ref{fig:nlSSE-results-at-50K}, magnon-polaron induced anomalies are also observed at the touching fields $H_{\rm TA,LA}$, which shift toward lower $H$ values by increasing the Bi amount [see also the magnified views for $H>0$ shown in Figs. \ref{fig:nlSSE-results-at-50K}(d), \ref{fig:nlSSE-results-at-50K}(e), and \ref{fig:nlSSE-results-at-50K}(f)].  
The $H_{\rm TA(LA)}$ values are evaluated as $\mu _0 H_{\rm TA} = 2.48~\textrm{T}$ ($\mu _0 H_{\rm LA} = 9.17~\textrm{T}$) for the Pt/YIG/Pt,  $2.13~\textrm{T}$ ($8.14~\textrm{T}$) for the Pt/Bi$_{0.5}$Y$_{2.5}$IG/Pt, and $2.02~\textrm{T}$ ($7.38~\textrm{T}$) for the  Pt/Bi$_{0.9}$Y$_{2.1}$IG/Pt, in good agreement with those estimated from the LSSE results. 
For the nlSSE, the magnon-polaron formation decreases rather than increases the ${\tilde S}$ signal at the touching fields $H_{\rm TA,LA}$, causing dip structures (note that the background nlSSE signal is opposite in sign compared to that for the LSSE\cite{Cornelissen2017PRB,Shan2018APL,Oyanagi2020AIPAdv}). 
In Ref. \onlinecite{Cornelissen2017PRB}, the dip feature is discussed in terms of the competition between a temperature-gradient induced magnon current, ${\bf J}_{\rm m}^{T} =- (\zeta/T) \nabla T$, from the injector to detector and a diffusive backflow, ${\bf J}_{\rm m}^{\mu} =- \sigma_{\rm m} \nabla \mu_{\rm m}$, induced by the gradient of the magnon chemical potential $\mu_{\rm m}$ [${\bf J}_{\rm m}^{T} \; || +{\bf \hat z}$ and ${\bf J}_{\rm m}^{\mu} \; || -{\bf \hat z}$ in the coordinate system shown in Fig. \ref{fig:LSSE-nlSSE}(b)], which are both increased by the hybridization with phonons when  $\tau^{-1}_{\rm mag} > \tau^{-1}_{\rm ph}$ \cite{Cornelissen2017PRB,Flebus2017PRB}. In particular, assuming that the magnon spin conductivity $\sigma_{\rm m}$ is more strongly increased than the bulk SSE coefficient $\zeta$ via the magnon-polaron formation, the increased backflow of magnons toward the injector direction causes the reduction of magnon accumulation beneath the Pt detector, which leads to the suppression of the nlSSE at the touching fields \cite{Cornelissen2017PRB}. \par
%
%
%
\begin{figure}[htb]
\begin{center}
\includegraphics[width=6.5cm]{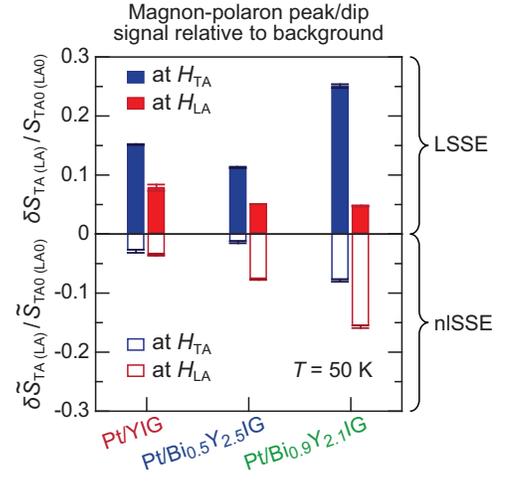}
\end{center}
\caption{
Comparison between the intensities of magnon-polaron LSSE-peak and nlSSE-dip
relative to the background, $\delta {S}_i/S_{i0}$ for the LSSE and $\delta {\tilde S}_i/{\tilde S}_{i0}$ for the nlSSE ($i = $ TA or LA), at the touching fields $H_{\mathrm{TA}}$ (blue bar chart) and $H_{\mathrm{LA}}$  (red bar chart) at $T=50~\textrm{K}$ for all the samples.  
}
\label{fig:MP-intensity-comparison}
\end{figure}
Through a careful look at the magnon-polaron anomalies in Fig. \ref{fig:nlSSE-results-at-50K}, we notice that the intensity of the anomaly at $H = H_{\rm LA}$ is larger than that at $H_{\rm TA}$ in the nlSSE for all the samples: $| \delta {\tilde S}_{\mathrm{TA}}/{\tilde S}_{\mathrm{TA0}} | < | \delta {\tilde S}_{\mathrm{LA}}/{\tilde S}_{\mathrm{LA0}} |$.  
The $\delta {\tilde S}_{\rm TA(LA)}/{\tilde S}_{\rm TA0(LA0)}$ values at $H_{\rm TA(LA)}$ are evaluated to be $-0.029$ ($-0.035$) for the Pt/YIG/Pt,  $-0.014$ ($-0.077$) for the Pt/Bi$_{0.5}$Y$_{2.5}$IG/Pt, and $-0.079$ ($-0.16$) for the  Pt/Bi$_{0.9}$Y$_{2.1}$IG/Pt at $T = 50~\textrm{K}$, which we summarize in Fig. \ref{fig:MP-intensity-comparison}.
The observed relationship for the nlSSE, $| \delta {\tilde S}_{\mathrm{TA}}/{\tilde S}_{\mathrm{TA0}} | < | \delta {\tilde S}_{\mathrm{LA}}/{\tilde S}_{\mathrm{LA0}} |$, holds in a wide $T$ range for $T \lesssim 100~\textrm{K}$, and is opposite to that for the LSSE results, where $\delta S_{\mathrm{TA}}/S_{\mathrm{TA0}} > \delta S_{\mathrm{LA}}/S_{\mathrm{LA0}}$ (see Fig. \ref{fig:MP-intensity-comparison} and also Figs. S1 and S2 in the Supplemental Material \cite{SM} for the $T$ dependence of the nlSSE). 
In the following, we discuss possible origins of the anisotropic feature.
One possible scenario is a spectral non-uniform magnon current $J_{\rm m}(\omega)$ \cite{Kikkawa2015PRB,Schmidt2021PRB} that may vary depending on the experimental configurations; the frequency-resolved current $J_{\rm m}(\omega_{\rm MTA})$ at $H_{\rm TA}$ may be larger (smaller) than  $J_{\rm m}(\omega_{\rm MLA})$ at $H_{\rm LA}$ for the longitudinal (nonlocal) configuration. 
In Refs. \onlinecite{Kikkawa2015PRB}, \onlinecite{Oyanagi2020AIPAdv} and \onlinecite{Jin2015PRB}, through high-$H$ dependence experiments and its comparison with theory, a spectral non-uniformity in a magnon current is suggested to be present in both LSSEs and nlSSEs due to the frequency-dependent  magnon scattering time \cite{Schmidt2021PRB,Streib2019PRB}.
Moreover, in the present case, as shown in Figs. \ref{fig:LSSE-results-at-50K} and \ref{fig:nlSSE-results-at-50K}, the high-$H$ response of the background SSEs in the longitudinal and nonlocal configurations differs with each other. This may be a signature that the frequency-dependent  $J_{\rm m}(\omega)$ contribution is different between the LSSE and nlSSE.  
Another possibility would be the anisotropy in magnon and magnon-polaron dispersion relations due to the nature of the dipole and magnetoelastic interaction \cite{Shen2015PRL,Flebus2017PRB}, which depend on the magnon propagation direction ${\bf J}_{\rm m}$ ($||~{\bf k}~||~\nabla T$) relative to the external magnetic field ${\bf H}$. 
In the LSSE, ${\bf J}_{\rm m}$ is essentially perpendicular to ${\bf H}$, while in the nlSSE parallel to ${\bf H}$ in the one-dimensional limit (see Fig. \ref{fig:LSSE-nlSSE}). 
Therefore, the LSSE and nlSSE may be affected by the anisotropic dispersion relations in a different manner.  
In fact, a solution of a Boltzmann transport equation shows that, for $\nabla T \perp {\bf H}$ (compatible with the LSSE), the magnon-polaron anomaly at $H=H_{\rm TA}$ is larger than that at $H_{\rm LA}$, while, for $\nabla T~||~{\bf H}$ (compatible with the nlSSE in the one-dimensional limit), the anomaly at $H_{\rm TA}$ is smaller than that at $H_{\rm LA}$ \cite{Flebus2017PRB}.
This is consistent with our experimental results, and suggests that the anisotropic nature of magnon and magnon-polaron dispersion relations may affect the SSEs. It is worth mentioning that the profile of the temperature gradient created by the local Pt heater in the nonlocal configuration may be affected by size effects \cite{Cornelissen2017PRB,Shan2016PRB,Shan2017PRB}, such as the thickness of the magnetic layer and the size of the local heater. Future systematic measurements of the Bi:YIG thickness, heater size, and injector-detector separation distance ($d$) dependencies may provide useful information to understand possible roles of the size effects in the anisotropic feature of magnon-polaron signals. Besides, the magnon-polaron formation can also affect the magnon and phonon thermal conductivities \cite{Flebus2017PRB}, which may modify the temperature gradient and resultant spin-current intensity at the onset field of magnon-polaron formation. The process, which has not been considered in analysis so far, would be important for further quantitative argument on the magnon-polaron anomalies in SSEs. \par 
%
\section{CONCLUSION} \label{sec:conclusion}
%
To summarize, we have prepared LPE-grown single crystalline Bi$_x$Y$_{3-x}$Fe$_{5}$O$_{12}$ (Bi$_x$Y$_{3-x}$IG; $x=0$, $0.5$, and $0.9$) films with Pt contact and measured SSEs in both the longitudinal and nonlocal configurations.
We observed two anomalous peaks in $H$-dependent LSSE signals at the onset fields $H_{\rm TA, LA}$ for the magnon-polaron formation at which the magnon and TA,LA-phonon branches in Bi$_x$Y$_{3-x}$IG touch with each other. The anomaly fields shift toward lower values by increasing the Bi amount $x$, which is attributed to the reduction of the sound velocities of Bi$_x$Y$_{3-x}$IG by the Bi substitution. The observed peak behavior at the wide temperature range from 3 to 300 K suggests that the scattering rate of magnons $\tau^{-1}_{\rm mag}$ is larger than that of phonons $\tau^{-1}_{\rm ph}$ in all our sample systems \cite{Flebus2017PRB}. 
In the nlSSE measurements, we found that the magnon-polaron formation suppresses rather than enhances the signal, causing dip structures at the touching fields $H_{\rm TA, LA}$ for the Bi$_x$Y$_{3-x}$IG films, consistent with the previous studies  \cite{Cornelissen2017PRB,Shan2018APL,Oyanagi2020AIPAdv}. We further found that the intensities of magnon-polaron anomalies in the nlSSE appear to be different from those in the LSSE, although the anomaly fields are almost identical for both the configurations.   
We anticipate that our detailed temperature and magnetic-field dependent SSE results provide useful information to further elucidate the physics of magnon-polaron SSEs. \par  
%
\section*{ACKNOWLEDGMENTS}
%
The authors thank S. Ito and Y. Murakami from Institute for Materials Research, Tohoku University, for performing transmission electron microscopy and electron probe microanalysis on our samples, respectively.
This work was supported by ERATO \textquotedblleft Spin Quantum Rectification Project\textquotedblright\ (No. JPMJER1402) and CREST (Nos. JPMJCR20C1 and JPMJCR20T2) from JST, Japan, 
Grant-in-Aid for Scientific Research (JP19H05600, JP20H02599, JP20K22476, JP21K14519, JP22K14584, and JP22K18686) and Grant-in-Aid for Transformative Research Areas (No. JP22H05114) from JSPS KAKENHI, Japan, 
NEC Corporation, and Institute for AI and Beyond of the University of Tokyo. 
Z.Q. acknowledges support from the National Natural Science Foundation of China (Grants No. 11874098 and No. 52171173). 
R.R. acknowledges support from the European Commission through the project 734187-SPICOLOST (H2020-MSCA-RISE-2016), the European Union's Horizon 2020 research and innovation program through the MSCA grant agreement SPEC-894006, Grant RYC 2019-026915-I funded by the MCIN/AEI/ 10.13039/501100011033 and by "ESF investing in your future", the Xunta de Galicia (ED431B 2021/013, Centro Singular de Investigación de Galicia Accreditation 2019-2022, ED431G 2019/03) and the European Union (European Regional Development Fund - ERDF). \par
%
%
%
%

\newpage 
\end{document}